\documentclass[a4paper,12pt]{nature2a}

\usepackage{latexsym}
\usepackage[dvipdfmx]{graphicx}
\usepackage{times}
\usepackage{color}
\usepackage{amsmath,amssymb,bm}
\usepackage{here}
\usepackage{multirow}

\usepackage[bf,labelsep=none,figurename=Figure\ , tablename=Table\ , singlelinecheck=true]{caption}
\usepackage[CaptionAfterwards]{fltpage}

\usepackage[square, super, numbers, merge, sort&compress]{natbib}
\makeatletter
\renewcommand{\section}{%
  \@startsection
    {section}%
    {1}%
    {\z@}%
    {0.1cm \@plus1ex \@minus .2ex}%
    {0.1cm}%
    {\normalfont\Large\bfseries}%
}%

\makeatother
\makeatletter
\renewcommand
{\subsection}{%
  \@startsection
    {subsection}%
    {2}%
    {\z@}%
    {0.1mm \@plus1ex \@minus .2ex}%
    {0.1mm}%
    {\normalfont\normalsize\bfseries}%
}%

\makeatother

\title{Interplay between Multipolar Order and Multipole-Induced Superconductivity in PrTi$_{2}$Al$_{20}$}

\author{Akito Sakai$^{1,2}$, Yosuke Matsumoto$^{2,3}$, Mingxuan Fu$^{1,2}${, Takachika Isomae$^{1,2}$}, Masaki Tsujimoto$^{2}$, Eoin O'Farrell$^{2}${, Daisuke Nishio-Hamane$^{2}$} and Satoru Nakatsuji$^{1,2,4-7*}$}

\begin{document}
\maketitle

\begin{affiliations}
\item Department of Physics, University of Tokyo, Hongo, Bunkyo-ku, Tokyo, 113-0033, Japan
\item Institute for Solid State Physics, University of Tokyo, Kashiwa, Chiba 277-8581, Japan 
\item Max Planck Institute for Solid State Research, Heisenbergstrasse 1, Stuttgart 70569, Germany 
%\item Center for Quantum Devices, Niels Bohr Institute, University of Copenhagen and Microsoft Quantum Lab Copenhagen, Universitetsparken 5, Copenhagen, 2100, Denmark 
\item Institute for Quantum Matter and Department of Physics and Astronomy, Johns Hopkins University, Baltimore, Maryland 21218, USA
\item Trans-Scale Quantum Science Institute, University of Tokyo, Bunkyo-ku, Tokyo 113-0033, Japan
\item Canadian Institute for Advanced Research (CIFAR), Toronto, Ontario M5G 1M1, Canada
\item CREST, Japan Science and Technology Agency (JST), 4-1-8 Honcho Kawaguchi, Saitama 332-0012, Japan 
\end{affiliations}

* email: satoru@phys.s.u-tokyo.ac.jp

\section*{Abstract}

\begin{abstract}
Multipolar moments entail a new route to tackle frontier problems in superconductivity (SC). A key progress in the search for multipolar SC is the discovery of Pr$Tr_2$Al$_{20}$ ($Tr =$ Ti, V), which possesses quadrupolar and octupolar but no magnetic dipolar moments. The Kondo entanglement of these multipolar moments with conduction electrons leads to exotic SC within the multipolar ordered phase, though the precise nature of the SC remains unexplored. We experimentally investigate the SC gap structure of SC in PrTi$_{2}$Al$_{20}$ and its La-doping evolution. Our results indicate deviations from a single $s$-wave gap, instead favoring nodal $d$-wave or multiple gaps. While the SC is robust against La dilution, the SC gap structure changes with minimal La doping, coinciding with a sharp change in the ferroquadrupolar (FQ) order. This suggests an intimate link between the quadrupolar order parameter and SC pairing, providing insight into the coexistence of SC with multipolar order.

% We found large Sommerfeld coefficient of 0.23 J/molK$^2$ in the normal state and its jump at $T_{\rm c}$ reaching $\Delta C_{4f}/T_{\rm c}\sim 0.11$ J/molK$^2$ , which are clear evidences of the bulk heavy fermion superconductivity. Furthermore, temperature dependence of $C/T$ and $B_{\rm c1}$ below $T_{\rm c}$ are found consistently reproduced by the two-band model. The contribution from the heavier band is rapidly suppressed by the small amount of La substitution, suggesting the quadrupole degree of freedom in the non-Kramers $\Gamma_3$ ground doublet, which becomes ill-defined due to the disorder, could be a key ingredient for the heavy fermion superconductivity in PrTi$_{2}$Al$_{20}$.
 \end{abstract}

\section*{Introduction}
With far-reaching impact on both fundamental research and technological innovations, the unconventional SC remains one of the most astonishing yet hardest problems to crack in quantum materials. The phase diagrams of various exotic superconductors display striking similarities despite their radically different parent materials; namely, the SC dome arises on the verge of ordered states entwining spin, orbital, and charge degrees of freedom (d.o.f) \cite{Keimer2015, Si2016, Onuki2004}. While early works point to spin fluctuations as the primary driver for the unconventional SC \cite{Toru2000, Monthoux2007}, the successive discoveries of orbital-driven nematicity in iron-based compounds \cite{Chu2010, Kasahara2012, Hosoi2016, Sprau2017}, copper oxides \cite{Ando2002,Kohsaka2007, Hinkov2008,Daou2010,Sato2017}, and rare-earth heavy fermions { \cite{Nakatsuji2008,Matsumoto2011,Ronning2017, Helm2020, Wu2020}} put forward orbital instability as a crucial ingredient for the pairing glue \cite{Maier2014, Lederer2015, Fradkin2010}. Understanding the role of orbital fluctuations on the superconducting properties may yield new knowledge that helps to identify the genes of exotic SC, and thereby, widening its landscape. 
 
The complex tangle of multiple d.o.f hinders a direct exploration of how the orbital instability affects SC properties \cite{Fernandes2014, Matsuura2017,Reiss2020}. The key to resolving such complication is a model system in which emergent electronic phenomena are governed solely by orbital d.o.f. An ideal material platform of this kind is the cubic 4$f$-electron system Pr$Tr_2$Al$_{20}$ ($Tr =$ Ti, V) that possesses a $\Gamma_3$ crystal electric field (CEF) ground state with quadrupolar and octupolar, but no magnetic dipolar moments \cite{akito}. The nonmagnetic ground-state doublet is well separated from the first-excited triplet with a CEF gap $\Delta_{\rm CEF} \sim 65$ K \cite{sato}, and therefore, the multipolar d.o.f dominates the low-temperature behavior. 

%{}{Among this family of materials, {PrTi$_2$Al$_{20}$} exhibits a ferroquadrupolar (FQ) order with order parameter $O_{20}$ at $T_{\rm Q}\sim 2.0$ K, followed by a SC transition at $T_{\rm c} \sim 0.2$ K, which is so far the highest $T_{\rm c}$ reported in pure multipolar systems \cite{akito, sato,Ito,nakanishi,Taniguchi2016,akitoSC}. }

{}{A unique characteristic of Pr$Tr_2$Al$_{20}$ ($Tr =$ Ti, V) is their strong hybridization between the local $4f$-multipolar moments and conduction ($c$) electrons that leads to enhanced quadrupolar Kondo effect and multipole-type RKKY coupling, as confirmed by various experimental probes\cite{akito,Matsunami,Tokunaga}. Among the known pure multipolar systems, PrTi$_2$Al$_{20}$ exhibits the highest quadrupolar ordering temperature $T_{\rm Q}\sim 2.0$ K and the highest superconducting transition $T_{\rm c} \sim 0.2$ K, which is directly associated with the strong $c$-$f$ hybridization.} Pressure-tuning of the $c$-$f$ hybridization strength in PrTi$_{2}$Al$_{20}$ renders a phase diagram that involves a SC dome lying inside the FQ ordered state. As the system approaches the boundary of the FQ order under $\sim 8$ GPa, the $T_{\rm c}$ value undergoes a five-fold increase, accompanied by a dramatic enhancement of the effective mass from $m^* \sim 16~m_0$ at ambient pressure to $\sim 110~m_0$ at $\sim 8$ GPa. {}{The coexistence of the SC dome with a pure FQ order indicates that the interaction between multipolar moments and conduction electrons is crucial for Cooper pairing. The pressure evolution of $T_{\rm c}$ and $m^*$ identifies a putative quantum critical point (QCP), near which quantum-critical quadrupolar (i.e., orbital) fluctuations substantially enhance the unconventional superconducting pairing mechanism\cite{Matsubayashi,*Matsubayashi2013}. }

{}{The experimental findings in PrTi$_{2}$Al$_{20}$ have sparked theoretical explorations into the impact of the multipolar Kondo coupling and multipolar order parameter on the nature of the SC state \cite{Nomoto2016, Sim2020, Kubo2020}.} The multipolar Kondo coupling generates an intimate spin-orbital entanglement of the conduction electrons in a multi-orbital system, promoting SC states characterized by higher-angular momentum Cooper pairs with $J > \frac{1}{2}$ \cite{patri2021SC} --- a promising route to intrinsic topological SC. \cite{Sato2017_review,Savary2017,Kim2018,Sim2020}
{}{Specific for cubic $\Gamma_3$ non-Kramers systems, nodal $d$-wave superconductivity is predicted to coexist with the $O_{20}$ quadrupolar order for an extended parameter space because they belong to the same irrep of symmetry\cite{Kubo2020}.  
Nonetheless, an in-depth experimental characterization of the gap symmetry of the multipolar SC and its evolution with non-thermal tuning is still lacking; such research could provide an invaluable guide for understanding the interplay between the long-range multipolar order and the coexisting SC. In the present study, we extensively investigate the superconducting properties of PrTi$_{2}$Al$_{20}$ and La-diluted compounds Pr$_{1-x}$La$_x$Ti$_2$Al$_{20}$ ($x\leq 1$). The ultralow-temperature specific heat and d.c. magnetization measurements on PrTi$_{2}$Al$_{20}$ represent the first thermodynamic characterization of SC coexisting with a pure multipolar order. The temperature dependence of the specific heat and the lower and upper critical fields, $B_{c1}$ and $B_{c2}$, reveal signatures that strongly deviate from those of isotropic $s$-wave SC but can be equally accounted for by single $d$-wave or multiple-gap structures. The La-doping evolution of the SC state and the FQ order indicates that the long-range FQ order in the clean limit plays a crucial role in shaping the SC gap structure.}

\section*{Results}

All measurements were carried out on high-quality PrTi$_2$Al$_{20}$ and La-doped PrTi$_2$Al$_{20}$ single crystals synthesized by the Al-self-flux method { with special care for producing the homogeneous mixture of Pr and La (Methods)}. {}{The La doping in the measured samples is confirmed to be homogeneous by scanning electron microscopy with energy dispersive X-ray analysis (SEM-EDX).} Details of sample characterization and experimental techniques are described in Methods. 

{}{The SC state of the undoped PrTi$_2$Al$_{20}$ displays behavior that is sharply different from that of a single-gap $s$-wave superconductor in various physical quantities: (1) The SC-induced specific heat anomaly shows a broad shoulder that cannot be described by the single-gap $s$-wave model (Fig. 1a); (2) The lower critical field $B_{c1}$ exhibits a linear temperature dependence without saturation down to the lowest measured temperature of 40 mK, a strong departure from the expected convex curvature expected for a single-gap $s$-wave pairing (Fig. 1b); (3) The upper critical field $B_{c2}$ shows an upturn curvature near $T_{c}$ instead of following the single-gap Werthamer-Helfand-Hohenberg (WHH)-like behavior (Fig. 2d), which is typical for a single anisotropic SC gap or a multigap SC state.\cite{Prohammer1990,Ando1999,Zhitomirsky2004,Mun2012}}

{}{In the following, we first explore the possible SC gap structure in PrTi$_2$Al$_{20}$ based on low-$T$ thermodynamic properties, focusing on the nodal $d$-wave pairing and multigap scenarios, which have been proposed to be relevant for Pr-based multipolar systems\cite{Kubo2020, patri2021SC}. Next, we examine the effect of La doping on the SC and FQ states.}

\subsection*{{}{Thermodynamic Characterization of the Superconducting State in PrTi$_2$Al$_{20}$}}

Figure 1 shows the specific heat divided by temperature, $C(T)/T$, of PrTi$_2$Al$_{20}$ at zero field and 10 mT. The pronounced peak in $C(T)/T$ at $T_{\rm Q}\sim$ 2.0 K marks a transition to the FQ ordered state, consistent with earlier reports \cite{akito}. Below $T_{\rm Q}$, $C/T$ follows an exponential decay function, namely, $C/T =\gamma + B \exp(-\Delta / T)$ (solid curve in Fig. 1), which yields a gap $\Delta = 2.4$ K and a Sommerfeld coefficient $\gamma=0.23$ (J/molK$^2$); this $\gamma$ is one order of magnitude larger than the reported value for the isostructural, non-4\textit f analog LaTi$_2$Al$_{20}$ \cite{Yamada2018}, which evidences the formation of heavy-fermion state most likely triggered by the quadrupolar Kondo effect in PrTi$_2$Al$_{20}$. On further cooling, a specific heat jump signals the onset of SC. The SC transition temperature $T_{\rm c} \sim 0.16$ K determined from the midpoint of the jump is slightly lower than the value $T_{\rm c} \sim 0.2$ K obtained from the previous resistivity and magnetic susceptibility measurements \cite{akitoSC}. The substantial $\gamma$ combined with the $C/T$-jump indicate bulk, heavy-fermion superconductivity in PrTi$_2$Al$_{20}$. 

The jump in $C/T$  at $T_{\rm c}$ vanishes at $B =$ 10 mT, as expected from the small upper critical field $B_{\rm c2}\sim$  6 mT (see Fig. 2d) \cite{akitoSC}. The electronic specific heat divided by $T$, $C_{e}/T$, is thereby obtained by subtracting the normal-state specific heat measured at 10 mT from the zero-field data, namely, $C_{e}/T=(C({\rm 0\ mT})-C({\rm 10\ mT}))/T +\gamma$ (inset of Fig. 1a); {}{the Sommerfeld term $\gamma=0.23$ (J/molK$^2$) is present in both the normal and SC states.} Such extraction of the $C_{e}/T$ eliminates Pr nuclear contribution that leads to the upturn in $C(T)/T $ below about 40 mK. {}{The resulting SC-induced specific heat jump defies description by the BCS formula $\Delta_{\rm BCS} (T) = 1.76k_{\rm B}T_{\rm c}\delta_{\rm BCS}(T)$, (see dotted black line in Fig. 1a, inset), in stark contrast to the non-4$f$ analog LaTi$_2$Al$_{20}$, which is a single-gap BCS superconductor\cite{Yamada2018}.} Moreover, the relative height of the jump $\Delta C_{\rm e}/\gamma T \sim 0.48$ is considerably smaller than the weak-coupling BCS prediction of 1.43 and the reported value of 1.26 in LaTi$_2$Al$_{20}$ \cite{Yamada2018}. {}{Given the high purity of the undoped sample with RRR $\sim 100$, it is unlikely that the observed broadening of the specific-heat jump is caused by distributed $T_{\rm c}$; rather, it could reflect the intrinsic SC gap structure \cite{MgB2, Lu2Fe3Si5, Kittaka2014, Sato2018}.
We analyze the temperature variation of $C_{e}/T$ in the SC state using two models with different gap structures: (1) single $d$-wave gap $\Delta (T) = \Delta_{0,d}\delta_{\rm BCS}(T)\cos 2\phi$; and (2) two $s$-wave gaps $\Delta_i (T) = \Delta_{0,i}\delta_{\rm BCS}(T)$, as shown in the inset of Fig. 1a. These two models describe the specific-heat jump equally well. Specifically, the best fit to the single $d$-wave model (green solid line) yields a SC gap size of $\Delta_{0,d}= 0.26$ K; the two-gap model (red solid line) gives two $s$-wave gaps $\Delta_{0,i=1} = 0.12$ K and $\Delta_{0,i=2} = 0.30$ K, with respective weights 65\% and 35\% of the total density of states (DOS). The model with a mixture of one $d$-wave and one $s$-wave gap (i.e., $d+s$ wave model) can also reproduce the data, while it demonstrates a dominant fraction of the $d$-wave component with a gap size nearly identical to that obtained from the single $d$-wave model (Supplementary Information, Fig. S4a).}

{}{To more extensively examine the SC gap structure in PrTi$_2$Al$_{20}$, we analyze the temperature dependence of the lower critical field, $B_{\rm c1}$ using the single-gap $d$-wave and two-gap $s$-wave models}. The isothermal equilibrium magnetization curves $M_{\rm eq} (B)$ for $T < T_{c}$ are shown in the inset of Fig. 1b (see Supplementary Information for the details of determining $M_{\rm eq}$). In the Meissner state, the $M_{\rm eq} (B)$ decreases linearly with a slope of $\sim -1$ (solid line in Fig. 1b, inset); deviation from the initial linear behavior occurs at $B_{\rm c1}$, signaling the entrance into the mixed state with penetration of vortices. {}{Again, the $B_{\rm c1}$ vs. $T$ obtained from $M_{\rm eq} (B)$ data are equally accounted for by the two models (Fig. 1b, main panel, Supplementary Information), with exactly the same set of fitting parameters as in the specific heat analysis. The excellent agreement between the analysis of the two thermodynamic properties further validates the single $d$-wave and two-gap scenarios for the multipolar SC in PrTi$_2$Al$_{20}$. Note that neither $C_{e}/T$ nor $B_{\rm c1}$ show any sign of two $T_c$ anomalies. If the two-gap SC scenario indeed holds for this system, this observation suggests non-negligible interband coupling in the SC state, which is supported by the two-band fitting results for $B_{\rm c2}$ vs. $T$ (Supplementary Information).\cite{Nicol2005}}

{}{The ultralow-temperature data below 20 mK are crucial for conclusively distinguishing these two scenarios. For instance, the presence of $d$-wave pairing would result in a power-law decrease in $C_{e}/T$ in the low-$T$ limit, whereas an exponential decay would indicate the isotropic $s$-wave gap. However, such a temperature window is essentially not accessible for thermodynamic probes. Thus, the precise pairing symmetry of the SC state remains an open question that awaits further experimental investigations, such as field-orientation dependence of specific heat and nuclear quadrupole spin-lattice relaxation.}

\subsection*{Effect of La doping}
{}{Next, we turn to the nonmagnetic La substitution effect in PrTi$_{2}$Al$_{20}$, which opens an effective route to clarify the interplay of the multipolar SC and the quadrupolar ordering. The essential impact of the La doping on the Pr, thereby (1) decreasing the intersite quadrupolar-quadrupolar interaction; (2) Acting as negative pressure, weakening the $c$-$f$ hybridization and thus the quadrupolar Kondo coupling; (3) Leading to local disorders that randomly split the non-Kramers doublet. Below, we present the experimental manifestations of these effects, focusing on the doping evolution of the SC and FQ states.} 

{}{The SC transition observed in the resistivity $\rho(T)$ and a.c. susceptibility $\chi' (T)$ is surprisingly robust against La doping (Fig. 2a, b). The transition temperature $T_c$ shows a marked decrease from the undoped value at $x = 0.03$} and then remains nearly constant for a wide doping range before eventually rising towards  $\sim 0.5$ K (namely, $T_{\rm c}$ of LaTi$_{2}$Al$_{20}$) in the dilute limit. {}{Such robustness of $T_c$ against La doping is also reported for other Pr-based systems hosting pure quadrupolar orders\cite{KMatsumoto2015}, suggesting that it might be a characteristic feature of quadrupole-mediated SC.} 

In striking contrast with the weak doping dependence of $T_{\rm c}$, the upper critical field $B_{c2}$ vs. $T$ curves obtained from the a.c. susceptibility measurements display a dramatic change in the low doping regime (Fig. 2d). The temperature dependence of $B_{c2}$ for the undoped sample shows an initial concave curvature at $B \sim 0$, {}{which is again well described by the two-gap $s$-wave model (see solid line in Fig. 2d). Nonetheless, this upturn near $T_{c}$ does not uniquely signal multigap SC but can also be attributed to single-gap $d$-wave pairing symmetry\cite{Prohammer1990,Ando1999}.} This feature is no longer detectable even with a tiny amount of La doping $x=0.03$ {(Fig. 2d)}. {}{In fact, the $B_{c2} (T)$ measured for the low La-doping levels ($x$ = 0.03, 0.11, and 0.22) are all well reproduced by the single-band WHH model (see dashed lines in Fig. 2d). This result indicates that the superconducting gap structure undergoes a significant change in the presence of slight La doping.} Moreover, with increasing $x$, the width of the SC transition $\Delta T_c$ narrows despite a tenfold increase in the normal state residual resistivity $\rho_{0}$ (Fig. 2b, c). {}{The $\rho_{0}$ value peaks at around $x = 0.6$ (Fig. 2c), indicating that atomic randomness predominantly affects the carrier scattering process, which would be expected to broaden the SC transition. Thus, the observed monotonic decrease of $\Delta T_c$ with increasing $x$ is unrelated to doping-induced spatial inhomogeneity but may instead serve as further evidence for the drastic change in the SC gap structure.} 

{}{Accompanying the modification in the SC gap structure, a closer look at the normal-state properties in the La-doped samples reveals a radical change of the FQ order in the low doping range.} Specifically, the gapped behavior below $T_{\rm Q}$ is rather fragile against La substitution. As shown in Fig. 3c, the exponential decay of $\rho (T)$ observed in the undoped PrTi$_2$Al$_{20}$ yields an anisotropy gap, $\Delta_{\rm AG} =$ 2.2 K (solid line), consistent with the estimation from the specific heat data (Fig. 1a). This gapped behavior is drastically suppressed even with a tiny amount of La substitution $x=0.03$, with the gap size shrinking to approximately half of the undoped value, as shown in Fig. 3a. {}{In the low doping regime, the quadrupolar-ordering-induced anomaly in $C_{4f}/T$ and $\rho (T)$ remains, yet shifting to lower temperatures with progressively diminishing magnitude as $x$ increases (Fig. 3d, main panel and Fig. S3b). These features indicate that the long-range FQ order transforms to short-range in the presence of slight La doping $x\lesssim 0.1$,} thereby turning the sharp FQ transition accompanied by spontaneous symmetry breaking into a crossover. {}{Altogether, both the nature of the SC state and the quadrupolar ordering change abruptly with minimal La doping, pointing to a tight link between the SC gap symmetry and the FQ order parameter.}

{}{Unlike the sudden changes in the SC gap structure and FQ order near the $x =0$ limit, the $c$-$f$ hybridization and the intersite (Pr-Pr) quadrupolar interaction decline gradually with increasing $x$, which are evident from the doping dependence of the lattice parameter and $T_Q$, respectively.} With increased $x$, the lattice parameter linearly increases, indicating that the La doping may smoothly weaken the $c$-$f$ hybridization and thus the quadrupolar Kondo effect (see Fig. S1a in Supplementary Information). {}{Indeed, the $A$ coefficient of the normal-state Fermi liquid (FL) resistivity, $\rho = \rho'_{0}+AT^2$ for $T_{Q} \lesssim T \lesssim $ 20 K, decreases quasi-linearly as $x$ increases (Fig. 3b and Fig. S3a in Supplementary Information).} Such doping dependence of $A$ corresponds to a gradual suppression of the effective mass ($m^* \sim \sqrt{A}$), which may reflect the smooth decline of the $c$-$f$ hybridization. Another clue for the reduced hybridization strength arises from the doping variation of the Schottky anomaly in the specific heat, as shown in the inset of Fig. 3d. In the undoped PrTi$_2$Al$_{20}$, the anomaly is substantially broader than the one derived from the CEF model due to the strong $c$-$f$ hybridization. At small doping levels ($x < 0.2$), the amplitude of the anomaly increases, contradicting the disorder effects that would cause continuous broadening of the anomaly. Such behavior is in line with the reduced $c$-$f$ hybridization strength; namely, the Pr-$4f$ electrons become more localized with increasing La doping. The CEF model describes the anomaly reasonably well, suggesting that the CEF-level schemes remain almost unchanged in the low doping range. {}{Furthermore, the $T_Q$ anomaly persists for a wide doping range and eventually disappears for $x > 0.7$ (Fig. 2a and Fig. 3d). The $T_Q$ value only slightly diminishes by less than $25\%$ from $x = 0$ to $x = 0.46$, suggesting that the intersite quadrupolar interaction is at play for a large doping window, while its sample-averaged strength declines gradually with increasing $x$.}

{}{Meanwhile, the disorder effect associated with the La doping may lead to random splitting of the ground-state non-Kramers doublet due to the loss of local cubic symmetry and the randomly distributed strain. This disorder effect might be responsible for the broadening of the Schottky anomaly observed at moderate doping levels $0.2< x< 0.5$ (Fig. 3d). For $x > 0.7$, the $T_Q$ anomaly is fully suppressed (Fig. 2a and Fig. 3d), indicating that quadrupolar degrees of freedom are no longer active due to the substantial splitting of the ground-state doublet.} This scenario is further supported by the reasonable match of the normal-state $C_{4f}/T$ observed for $x = 0.73$ with that predicted for a random two-level system (dashed line in Fig. 3d) \cite{RTL1, RTL2}.

\section*{Discussion}

The SC pairing symmetry ties in with the order parameter of the coexisting long-range FQ order, and both alter dramatically at $x\sim 0$. {}{We first suggest the potential mechanism responsible for the modification of the SC gap structure. In undoped PrTi$_2$Al$_{20}$, the quadrupolar Kondo interaction yields Fermi surface (FS) sheets with heavy effective mass\cite{KuboFS2020}. The formation of long-range FQ order causes local tetragonal distortion and changes in the heavy-mass FS sheets; in this way, the FQ order parameter influences the gap symmetry of the multipolar SC\cite{Sim2020, patri2021SC}. With a small amount of La doping, the crossover from long-range to short-range FQ order occurs and removes the spontaneous symmetry breaking. As a result, the structural and Fermi surface distortions driven by quadrupolar ordering no longer take place, thereby destabilizing the associated SC gap.} 

{}{Given that both the single-gap $d$-wave and two-gap models effectively account for the SC properties observed in PrTi$_2$Al$_{20}$, we propose two possibilities for the doping-induced SC gap change: (1) The $d$-wave gap nodes vanish, turning into an $s$-wave SC gap as the FQ order parameter becomes ill-defined. This $s$-wave gap persists in the short-range FQ regime and smoothly evolves into the BCS $s$-wave SC in LaTi$_2$Al$_{20}$; (2) In the multigap scenario, the SC gap stemming from the heavy-mass FS sheets is markedly suppressed with a small amount of La substitution, suggesting that the heavy-mass FS contribution to the SC state is highly sensitive to doping. The remaining SC gap exhibits $s$-wave symmetry and arises from the light-mass FS sheets nearly identical to those observed in LaTi$_2$Al$_{20}$ (i.e., the $\alpha$, $\delta$, $\epsilon$, and $D$ branches with light carrier mass $m^{*} \sim$  1.25 - 2.36 $m_0$) \cite{KuboFS2020}. In the dilute limit, the quasi-linear variation of $T_c$ is consistent with the instability of the BCS pairing in the presence of magnetic Pr substitution. The proposed scenario (2) resembles the multiband SC reported in La-doped PrOs$_4$Sb$_{12}$\cite{Yogi2006}. Nonetheless, there is a crucial difference between PrTi$_2$Al$_{20}$ and PrOs$_4$Sb$_{12}$: the CEF gap in PrTi$_2$Al$_{20}$ is nearly an order of magnitude larger than that in PrOs$_4$Sb$_{12}$, making the magnetic scattering arising from the first-excited CEF state irrelevant to the observed heavy-fermion SC. In other words, the SC behavior observed in PrTi$_2$Al$_{20}$ is purely associated with multipolar moments of the non-Kramers ground doublet.}

The linear-in-$x$ increase of the lattice parameter suggests that the La dilution acts as negative pressure (Fig. S1a in Supplementary Information). {}{This experimental feature lays the foundation for constructing the effective pressure phase diagram (Fig. 4), which clearly demonstrates the mild doping variation of $T_Q$ and the robust $T_c$ that stretches into the very dilute limit.} Notably, this phase diagram is manifestly different from that of antiferromagnetic heavy-fermion superconductors, in which SC is typically confined near the border of the long-range magnetic order and is sensitive to chemical doping \cite{Nakatsuji2002}. 

{}{The observed weak doping dependence of $T_Q$ contrasts sharply with that reported in other Pr-based compounds hosting nonmagnetic $\Gamma_3$ ground-state doublet and quadrupolar order, such as Pr$_{1-x}$La$_x$Ir$_2$Zn$_{20}$ and Pr$_{1-x}$La$_x$Pb$_{3}$ in which $T_Q$ drops rapidly with increasing La doping and is completely suppressed for $x < 0.1$\cite{Kawae2001,KMatsumoto2015} Meanwhile, owing to the strong $c$-$f$ hybridization and consequently the increased RKKY coupling among quadrupolar moments, the undoped PrTi$_2$Al$_{20}$ possess nearly an order of magnitude larger $T_Q$ than PrIr$_2$Zn$_{20}$ and PrPb$_{3}$. This comparison suggests a connection between the $T_Q$ value at $x = 0$ and the sensitivity of $T_Q$ to doping.} {Moreover, fitting the specific heat curve obtained for $x = 0.73$ to the random two-level model yields a maximum energy splitting $E_0\sim 2.4$ K of the ground-state doublet (Fig. 3d), which is comparable to $T_Q\sim 2$ K at $x = 0$. From this, we can infer that the intersite quadrupolar interaction dominates the disorder-induced two-level splitting across a wide $x$ range until the maximum ground-state doublet splitting reaches the energy scale of the long-range FQ order, where a substantial suppression of $T_Q$ takes place.}

{}{Another intriguing feature revealed in the phase diagram is the distinct behavior of $T_c$ near the FQ phase boundary under applied pressure and chemical doping (Fig. 4). With applied pressure, the $c$-$f$ hybridization increases, leading to enhanced quadrupolar Kondo interaction that drives the suppression of FQ order accompanied by a pronounced upturn of $T_c$\cite{Matsubayashi}. By contrast, near the verge of the FQ phase on the doping side, $T_c$ remains nearly constant without noticeable enhancement. This $T_c$ behavior might result from the combined effects of reduced $c$-$f$ hybridization and disorder-induced ground-state doublet splitting, both weakening the quadrupolar Kondo coupling.}

{}{To conclude, we reveal thermodynamic signatures that indicate nodal $d$-wave symmetry or a multigap structure for the SC in PrTi$_{2}$Al$_{20}$.} Although the SC transition extends over a wide La doping range, {}{the SC gap structure and the nature of the FQ order are concomitantly modified with a small amount of La doping, despite the gradual doping variation of the intersite quadrupolar interaction and $c$-$f$ hybridization. This finding highlights the essential role of the quadrupolar order parameter in shaping the SC pairing structure.} The investigation of multipole-induced SC in the model material Pr$Tr_2$Al$_{20}$ ($Tr =$ Ti, V) will help to unravel a unifying mechanism stringing orbital d.o.f and unconventional SC in many different families of materials with distinct parent phases.

\newpage
\section*{Methods}
Single crystals of {undoped PrTi$_2$Al$_{20}$ and LaTi$_2$Al$_{20}$}  were synthesized by the Al-self-flux method after mixing element Pr and La by arc melting, following the same procedures described in our previous publication \cite{Kangas}. The PrTi$_2$Al$_{20}$ single crystals used for the specific heat and magnetization measurements were selected from the best batch with typical residual resistivity ratio (RRR $=\rho(300 {\rm K})/\rho(0.4 {\rm K})$) about 100. {To prepare homogeneous La-doped single crystals Pr$_{1-x}$La$_x$Ti$_2$Al$_{20}$, we first synthesized molten button of 50 at\% Pr and 50 at\% La using mono-arc furnace. Then, half of the molten button was used for synthesizing single crystals of Pr$_{1-x}$La$_x$Ti$_2$Al$_{20}$ ($x\sim 0.5$) by flux method. The remaining half was cut into two and mixed with the same mole number of Pr and La, respectively, by mono-arc furnace. This step results in $x\sim 0.75$ and $x\sim 0.25$ molten buttons that were used for subsequent single crystal growth of the above-mentioned $x$ and the next-step doping process for different $x$. This procedure is repeated to obtain single crystals of different La concentrations.}
The chemical composition $x$ of Pr$_{1-x}$La$_x$Ti$_2$Al$_{20}$ was determined by scanning electron microscopy-energy dispersive X-ray analysis (SEM-EDX). {Homogeneous La doping is confirmed by elemental mappings and line profiles of elemental distribution (Fig. S1 b, c).} The electrical resistivity was measured by the standard four-probe method. 
The specific heat was measured by the quasi-adiabatic thermal relaxation method using a commercial setup (PPMS) at 0.4 K $< T< $10 K; measurements at lower temperatures (0.03 K $< T< $1 K) are done with a homemade cell installed in a $^3$He-$^4$He dilution refrigerator\cite{Matsumoto2018}.  
The dc-susceptibility was measured by a homemade magnetometer comprising a commercial SQUID sensor (Tristan Technologies) installed in the $^3$He-$^4$He dilution refrigerator.
The ac-susceptibility was measured by a mutual inductance method under an ac field of $\sim 50$ mOe. A piece of pure aluminum shaped in the same geometry and is similar in size as  Pr$_{1-x}$La$_x$Ti$_2$Al$_{20}$ were placed inside the canceling coil as a reference. In both cases, the dc magnetic field was applied by a homemade superconducting magnet with a Nb superconducting shield, thereby preventing the magnetic flux from outside. The superconducting diamagnetic response of Pr$_{1-x}$La$_x$Ti$_2$Al$_{20}$ is nearly identical in magnitude to that of the aluminum reference, consistent with the bulk nature of the superconductivity in Pr$_{1-x}$La$_x$Ti$_2$Al$_{20}$. 

\section*{Data Availability}
{The data that support the findings of this study are provided in the Supplementary Source Data file. Additional raw data related to this study are available from the corresponding author upon reasonable request. }

\section*{References}

%\bibliography{bibfile_PrLaTi2Al20}

\section*{Acknowledgements}
 We thank T. Sakakibara, K. Ueda, M. Takigawa, and Y. B. Kim for insightful discussions. This work was partially supported by JST-Mirai Program (JPMJMI20A1), JST-ASPIRE (JPMJAP2317) and JSPS-KAKENHI (JP23K03298). The work at the Institute for Quantum Matter, an Energy Frontier Research Center was funded by DOE, Office of Science, Basic Energy Sciences under Award \# DE-SC0024469. M.F. acknowledges support from the Japan Society for the Promotion of Science Postdoctoral Fellowship for Research in Japan (Standard). M.T. was supported by Japan Society for the Promotion of Science through Program for Leading Graduate Schools (MERIT).

\section*{Author Contributions}
S.N. conceived the project.
A.S., M.T. and S.N. synthesized the single crystals and prepared the samples for measurements.
A.S., Y.M. {T.I.} and E.O. carried out the transport, specific heat, and magnetization measurements and analyzed the data.
A.S. performed chemical analyses. {D.N-H. performed element mapping.}
A.S., M.F., Y.M., and S.N. wrote the paper. 
All authors discussed the results and commented on the manuscript.

\subsection*{Corresponding author}
Satoru Nakatsuji (email: satoru@phys.s.u-tokyo.ac.jp)

\section*{Inclusion and Ethics}
\subsection*{Competing Interests }
The authors declare no competing interests.

\newpage

\begin{figure}[t]
\centering 
\includegraphics[keepaspectratio, width=14.5 cm]{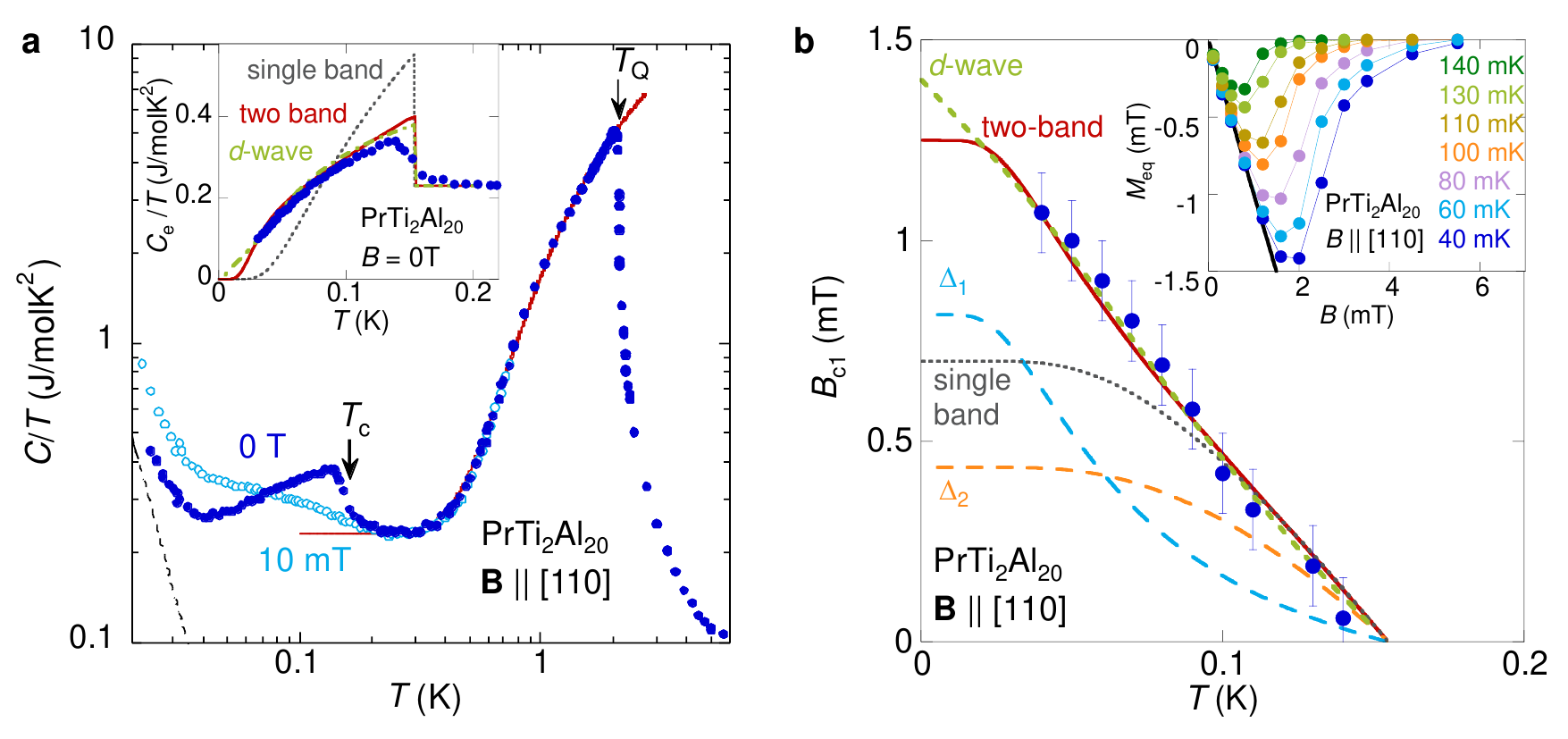}
\caption{{\bf $|$ {Multipole-driven} superconductivity in PrTi$_2$Al$_{20}$.} {\bf a,} Temperature dependence of the specific heat divided by temperature $C(T)/T$ of PrTi$_2$Al$_{20}$ at the field of $B =$ 0 T (solid) and 10 mT applied along the [110] direction (open). The solid line represents the fit $C(T)/T =\gamma + B \exp(-\Delta/ T)$, providing an anisotropy gap $\Delta = 2.4$ K and a Sommerfeld coefficient $\gamma=0.23$ (J/molK$^2$). {The dashed line represents the nuclear contribution (Supplementary Information).}
Inset: The electronic specific heat divided by temperature, {$C_{e}/T=(C({\rm 0\ mT})-C({\rm 10\ mT}))/T +\gamma$}, which is fitted to the single-band (dotted line) and two-band (solid line) models within the BCS framework{, and single $d$-wave model (dash-dotted line)}.  {\bf b,} $T$ dependence of the lower critical field $B_{\rm c1}$. The dotted and solid lines represent the single- and two-band fit, respectively. The two-band fit is generated using {exactly} the same set of parameters as for the specific heat data, and the dashed lines show the contribution from each band. Inset: The isothermal equilibrium magnetization $M_{\rm eq} (B)$ measured in $\mathbf{B}\parallel [110]$ at various $T$s under zero-field-cooled condition. Deviation from the initial linear behavior (solid line) marks the $B_{\rm c1}$, {and the error bars in the main panel are obtained by the linear fit}. }\label{fig1}
\end{figure}

\newpage

\begin{figure}[t]
\begin{center}
\includegraphics[keepaspectratio, width=16.3 cm]{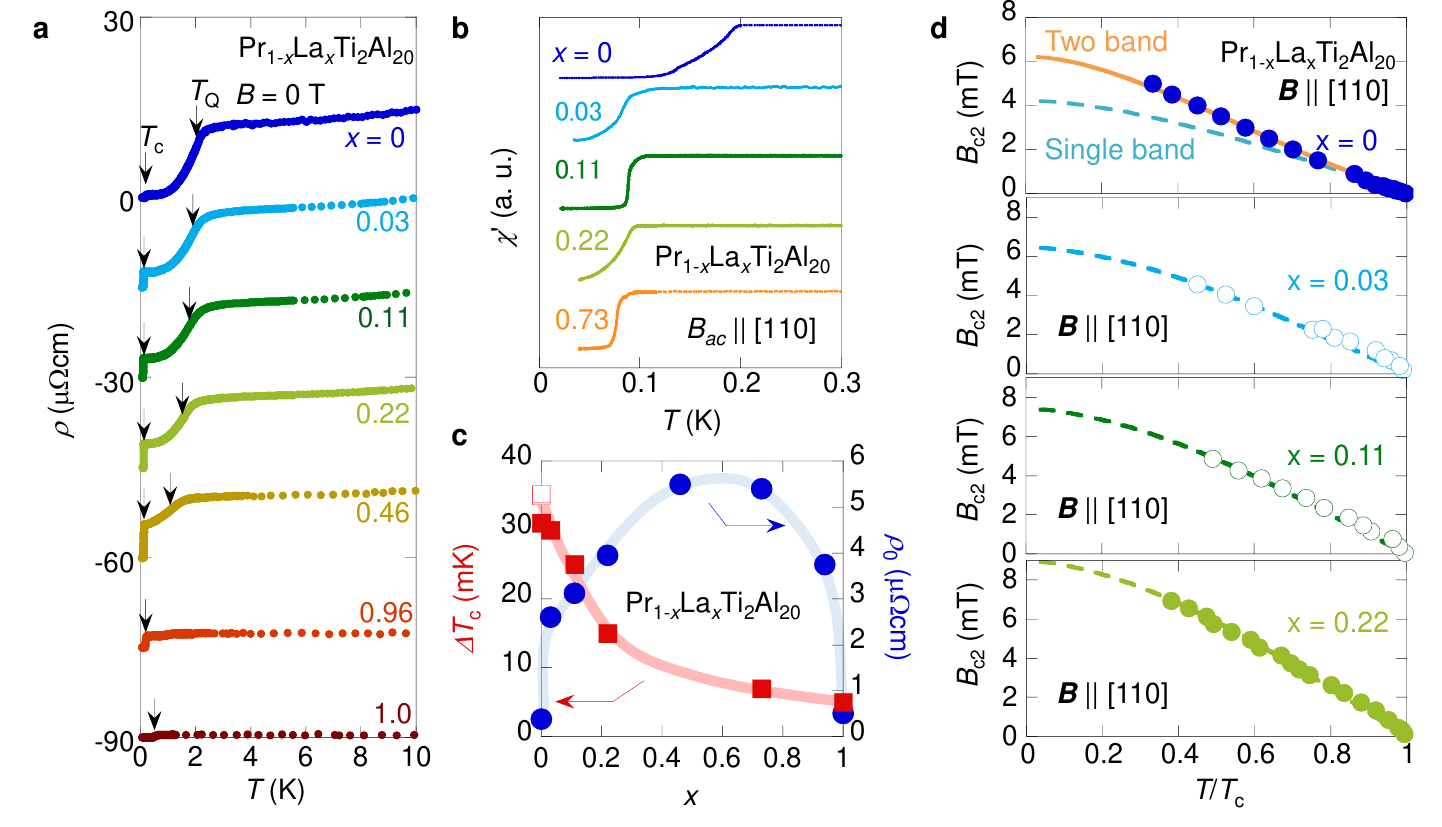}
\caption{{\bf $|$ Superconducting properties of Pr$_{1-x}$La$_X$Ti$_2$Al$_{20}$}  {\bf a,} The zero-field temperature dependence of the resistivity $\rho(T)$ for Pr$_{1-x}$La$_x$Ti$_2$Al$_{20}$ at various doping level $x$. The curves are shifted vertically for clarity. The arrows mark the ferroquadrupolar ordering temperature $T_{\rm Q}$ and the superconducting transition temperature $T_{\rm c}$. % The $4f$-electron contribution (dashed lines) forms a maximum at $T_{\rm peak}\sim 60$K, representing the crystal field splitting $\Delta_{\rm CEF}$ between the first excited triplet and the nonmagnetic ground-state doublet (see main text). 
{\bf b,} The real part of the ac susceptibility $\chi'$ as a function of $T$ at various La content $x$. The ac magnetic field is about 5 $\mu$T applied along the [110] direction. {\bf c,} Doping dependence of the superconducting transition width $\Delta T_{c}\equiv T_{c}-T_{\rm half}$ determined from the $T$-dependence of ac $\chi'$ {($\blacksquare$, left) and dc $\chi$ at $B=3$ mT ($\square$, left), and the residual resistivity $\rho_{0}$ ($\bullet$, right). Here, $T_{\rm half}$ is defined as $\chi(T_{\rm half})=(\chi(T_c)-\chi(T\sim 40 {\rm mK}))/2$}. {The solid lines are the guides to the eyes.}
{\bf d,} The upper critical field $B_{\rm c2}$ for Pr$_{1-x}$La$_x$Ti$_2$Al$_{20}$ ($x=0$, {0.03, 0.11,} 0.22). The {broken and} solid curves represent the fits by the {single-band and two-band} WHH model{, respectively (Supplementary Information)}.
}\label{fig2} 
\end{center}
\end{figure}

%(b) $T^2$ dependence of $\rho(T)$. (c) Semi-log plot of $\rho(T)-\rho_0$ vs $1/T$, where $\rho_0$ is a residual resistivity.}

\newpage

\begin{figure}[t]

\includegraphics[keepaspectratio, width=17cm]{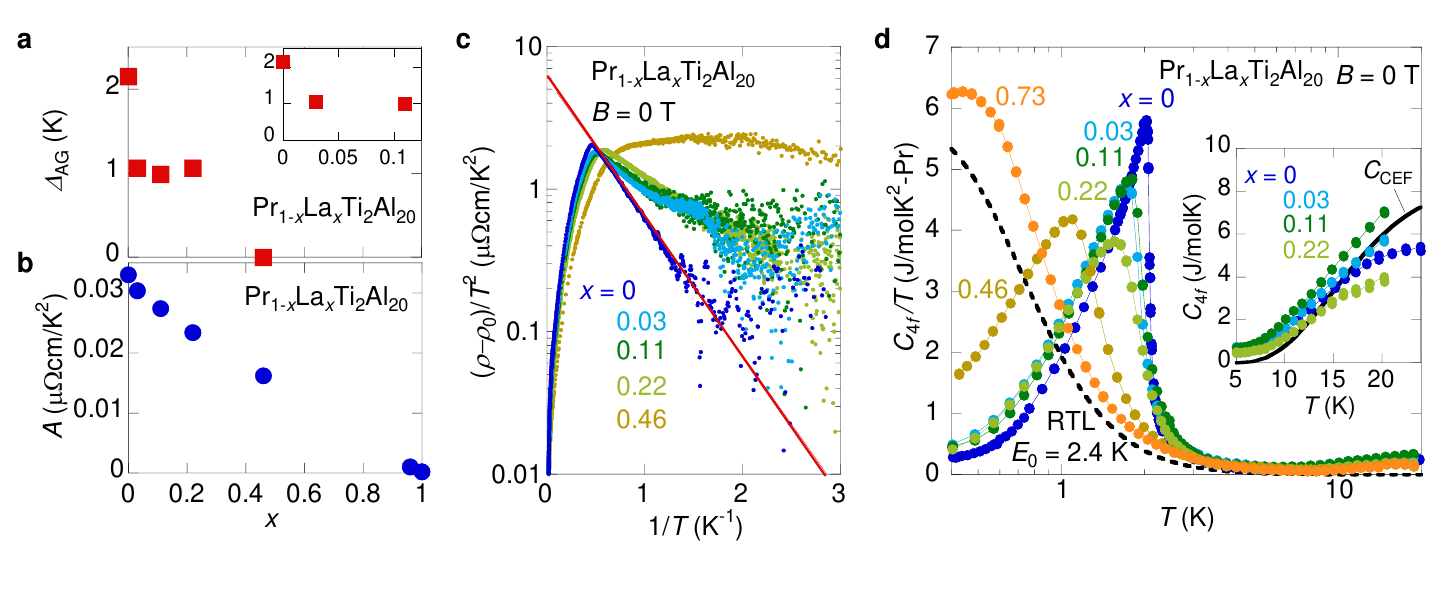}
\caption{{\bf $|$ Suppression of long-range quadrupole order in Pr$_{1-x}$La$_X$Ti$_2$Al$_{20}$.} {\bf a, b,} Doping dependence of the anisotropy gap $\Delta_{\rm AG}$ (a) and the $A$ coefficient (b). $\Delta_{\rm AG}$ and $A$ are determined by fitting the resistivity to $(\rho(T)-\rho_0)/T^2 \propto \exp (-\Delta_{\rm AG} /T)$ below $T_{Q}$ and $\rho(T) = AT^{2}+\rho'_{0}$ above $T_{Q}$, respectively. 
%Note that the residual resistivity $\rho'_{0}$ estimated from the FL fit is different from the actual $\rho_{0}$ estimated from the exponential fit shown in (b). 
{Inset: A zoomed plot of $\Delta_{\rm AG}$ vs. $x$ around $x\sim 0$.} 
{\bf c,} Semi-log plot of $(\rho(T)-\rho_0)/T^{2}$ vs $1/T$. The solid line represents the exponential fit $(\rho(T)-\rho_0)/T^2 \propto \exp (-\Delta_{\rm AG} /T)$ for $x = 0$, which gives $\Delta_{\rm AG} = 2.2$ K, consistent with the value estimated from the specific heat data. {\bf d,} The $4f$ contribution to the specific heat, $C_{4f}/T$, in the normal state. The dashed line in the main panel represents the prediction based on the random two-level (RTL) model $\int_0^{E_0} (1/E_0)(E/k_{\rm B}T)^2e^{-E/k_{\rm B}T}/(1+e^{-E/k_{\rm B}T})^2 dE$, where $E_0 =$ 2.4 K represents the cutoff of the energy splitting\cite{RTL1,RTL2}. Inset: A zoom plot of $C_{4f}$ vs. $T$ at high-$T$ region, where $C_{4f}$ shows a Schottky anomaly due to the CEF effect. The solid line represents the calculated specific heat by assuming the CEF parameters obtained from the neutron scattering \cite{sato}.
%Inset: A zoom plot of the high-$T$ region, where the $C_{4f}/T$ shows a Schottky anomaly due to the CEF effect. The solid line represents the calculated Schottky anomaly using the two-level model, by assuming that the first CEF excited state is located at 60 K above the ground-state doublet.
}
\label{fig3}

\end{figure}

\newpage
\begin{figure}[t]
\begin{center}
\includegraphics[keepaspectratio, width=16cm]{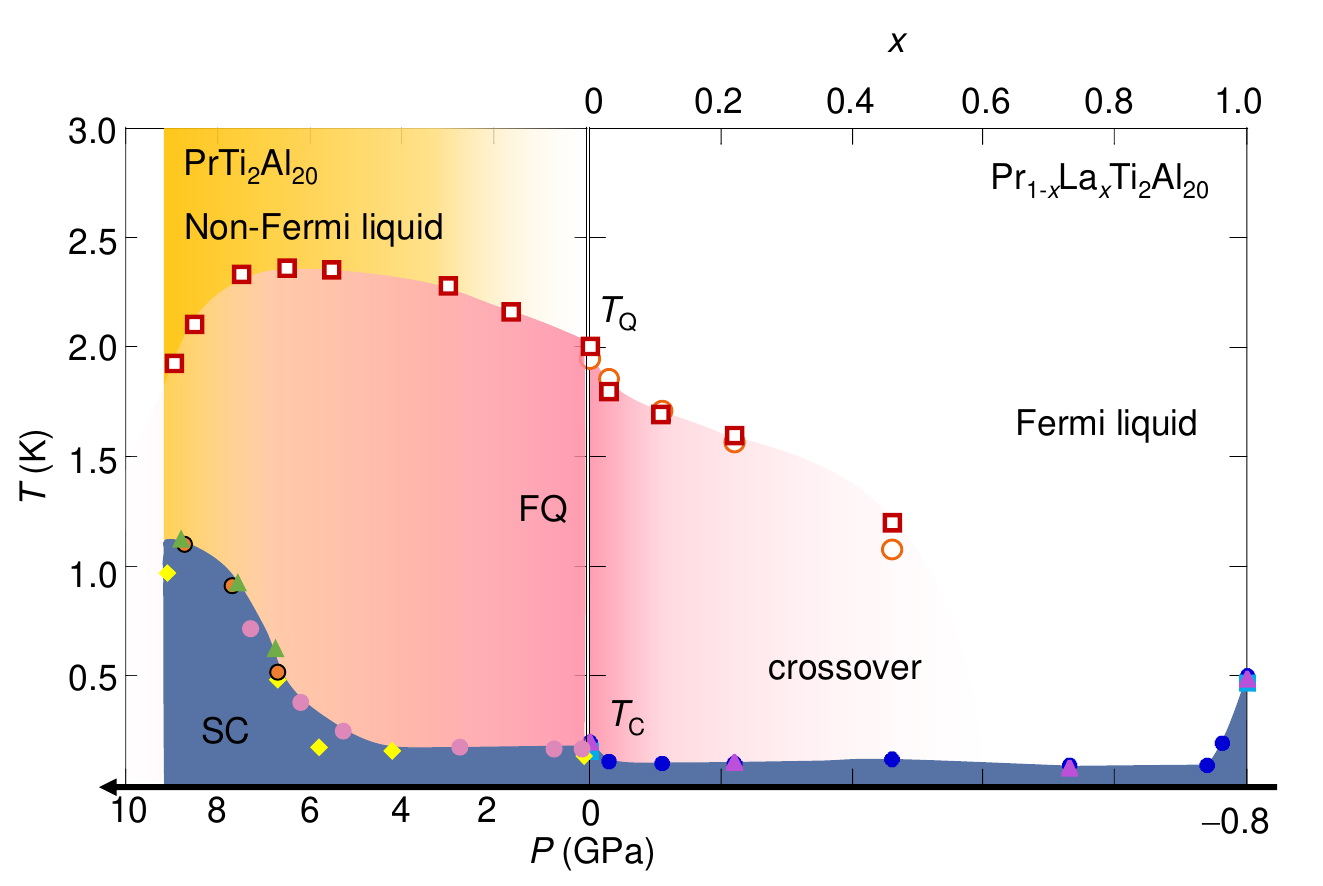}
\caption{{\bf $|$ Pressure/doping-temperature phase diagram for { the quadrupole Kondo lattice} PrTi$_2$Al$_{20}$.}  The FQ transition temperature $T_{Q}$ are determined from the temperature dependence of the specific heat (open squares) and resistivity {derivative (open circles, Supplementary Information, Fig. S3b)} for the La-doped samples; the superconducting transition temperature $T_{c}$ are deduced from the resistivity (filled circles), specific heat (filled squares), and the ac magnetic susceptibility (filled triangles) measurements. The data points under hydrostatic pressure are extracted from Ref. \cite{Matsubayashi, *Matsubayashi2013}. The La dilution causes a linear increase of the lattice parameter (see Fig. S1a in Supplementary Information), and therefore generates an effective negative pressure that can be tuned systematically with the doping amount $x$.}\label{fig4}
\end{center}
\end{figure}

\end{document}